\newcommand{\CLASSINPUTtoptextmargin}{2cm}
\newcommand{\CLASSINPUTbottomtextmargin}{2.5cm}
\documentclass[conference]{IEEEtran}
\IEEEoverridecommandlockouts

\usepackage[dvipdfmx]{graphicx}
\usepackage{lipsum,graphicx,subcaption}
\usepackage{float}
\usepackage{cite}
\usepackage{amsmath,amssymb,amsfonts,amsbsy}
\usepackage{algorithmic}
\usepackage{textcomp}
\usepackage{algorithmic,algorithm}
\usepackage{url}
\usepackage{color}
\usepackage[11pt]{moresize}

\def\BibTeX{{\rm B\kern-.05em{\sc i\kern-.025em b}\kern-.08em
    T\kern-.1667em\lower.7ex\hbox{E}\kern-.125emX}}

\newcommand{\rmnum}[1]{\romannumeral #1}
\usepackage{mathtools}
\DeclarePairedDelimiter\norm{\lVert}{\rVert}
\usepackage{amsthm}
\theoremstyle{definition}

\newtheorem{lemma}{Lemma}

\usepackage{mathabx}
\usepackage{optidef}
\usepackage{enumitem}
\usepackage{tabulary}
\usepackage{booktabs}
\usepackage{multirow}
\newcommand*{\Scale}[2][4]{\scalebox{#1}{$#2$}}

\begin{document}
\title{Lens-based Millimeter Wave Reconfigurable Antenna NOMA}

\author{
    \IEEEauthorblockN{Mojtaba~Ahmadi~Almasi\IEEEauthorrefmark{1}, Roohollah~Amiri\IEEEauthorrefmark{1}, Mojtaba~Vaezi\IEEEauthorrefmark{2}, and  Hani~Mehrpouyan\IEEEauthorrefmark{1}}\\
    \IEEEauthorblockA{\IEEEauthorrefmark{1}{Department of Electrical and Computer Engineering, Boise State University, Idaho, USA}}
    \IEEEauthorblockA{\IEEEauthorrefmark{2}{Department of Electrical and Computer Engineering, Villanova University, PA, USA}}
    \IEEEauthorblockA{Email:~{\{mojtabaahmadialm, roohollahamiri, hanimehrpouyan\}}@boisestate.edu, mvaezi@villanova.edu}
}
%


\markboth{}%
{Shell \MakeLowercase{\textit{et al.}}: Bare Demo of IEEEtran.cls for IEEE Journals}
\maketitle
\begin{abstract}
This paper proposes a new multiple access technique based on the millimeter wave lens-based reconfigurable antenna systems. In particular, to support a large number of groups of users with different angles of departures (AoDs), we integrate recently proposed reconfigurable antenna multiple access (RAMA) into non-orthogonal multiple access (NOMA). The proposed technique, named reconfigurable antenna
NOMA (RA-NOMA), divides the users with respect to their AoDs and channel gains. Users with different AoDs and comparable channel gains are served via RAMA while users with the same AoDs but different channel gains are served via NOMA. This technique results in the independence of the number of radio frequency chains from the number of NOMA groups. Further, we derive the feasibility conditions and show that the power allocation for RA-NOMA is a convex problem. We then derive the maximum achievable sum-rate of RA-NOMA. Simulation results show that RA-NOMA outperforms conventional orthogonal multiple access (OMA) as well as the combination of RAMA with the OMA techniques.

\end{abstract}
\section{Introduction}
Global mobile data traffic is projected to grow exponentially in the coming several years.  A similar trend is foreseen for the number of users. Specifically, fifth generation (5G) networks are envisioned to serve more than one million users per $km^2$~\cite{zhang2016fronthauling}. To meet these demands, millimeter wave (mmWave) communications and non-orthogonal multiple access (NOMA) have emerged as two key technologies.

The large unused spectrum available at the mmWave band (30-300 GHz) offers a great potential for the transmission of a substantial volume of data. Short wavelength of the mmWave band allows for employing a huge number of antenna elements at small space. To exploit this unique property, several designs have been proposed for mmWave systems. In one design, each antenna element is connected to one radio frequency (RF) chain which is called \textit{digital beamforming} system. The system achieves a high throughput, however, is costly both in terms of hardware complexity and power consumption. To support multiple streams while keeping the hardware complexity low, \textit{hybrid beamforming} systems are proposed~\cite{el2014spatially}. In this design, each RF chain is connected to the antennas via phase-shifters (PSs). In another design, the antennas are positioned on the surface of a lens and associated with RF chains via a switch network. Such a  system is termed \textit{beamspace} multiple-input multiple-output (MIMO)~\cite{r9}. The above systems mostly focus on improving the spectral efficiency in the mmWave spectrum, however, they fail to overcome severe path loss and shadowing. To this aim, recently, the concept of \textit{lens-based reconfigurable antenna MIMO} (RA-MIMO) has been introduced for mmWave communications~\cite{r19,almasi2018new}. In lens-based antennas, each RF chain is connected to just one reconfigurable antenna. Further, each reconfigurable antenna is able to steer multiple independent beams. 

NOMA is another enabling technology for 5G. NOMA in the power domain uses superposition coding at the transmitter and successive interference cancellation at the receiver to simultaneously serve multiple users and enhance spectral efficiency in multi-user scenarios~\cite{saito2013system, vaezi2018multiple}.
In mmWave communications, due to high path loss, users with different locations experience considerably different channel gains. This implies that mmWave systems better suit power domain NOMA which offers better spectral efficiency compared to that of orthogonal multiple access (OMA) techniques. In light of this, several groups have studied the integration of NOMA in mmWave systems, i.e., mmWave-NOMA~\cite{cui2018optimal,art_Moj_TCOM}. However, the major limitation with the current mmWave-NOMA technique is that the number of RF chains is the same as the number of NOMA groups. This limits the number of simultaneously supported user groups.

In our previous work~\cite{almasi2018reconfigurable}, we have shown that the rate performance of a multiuser system with a lens-based reconfigurable antenna~\cite{r19} is degraded when NOMA is applied. To tackle this issue, a new technique named reconfigurable antenna multiple access (RAMA) was proposed in~\cite{almasi2018reconfigurable}. Although RAMA achieves a higher sum-rate than both OMA and NOMA, its operation is limited to only one user per beam. To overcome this limit, we propose a new multiuser technique which integrates RAMA and NOMA, and is referred to as RA-NOMA. In RA-NOMA, each antenna element is connected to all RF chains. Users are grouped into two different sets based on the angle of departures (AoDs) and the users' channel gain. The users with different AoDs and comparable channel gains are supported via RAMA while the users with the same AoDs but different channel gains are served via NOMA. This technique introduces a new degree of freedom for grouping the users without adding RF chains. We show that, the multi-beam NOMA power allocation problem can be transformed into a convex problem over one beam. Further, the feasibility conditions of the power allocation problem are assessed. 
Numerical results reveal that the proposed design outperforms OMA and the combination of RAMA with conventional OMA techniques in terms of sum-rate.

The paper is organized as follows. In Section~\ref{sec:RAMA}, a brief background on RAMA is provided. Section~\ref{sec:RA_NOMA} proposes the RA-NOMA technique. Section~\ref{sec:power} presents the power allocation in the downlink of RA-NOMA. In Section~\ref{sec:sim}, simulation results are presented. Finally, Section~\ref{sec:conclusion} concludes the paper.

\textbf{Notations:} Hereafter, $j = \sqrt{-1}$. Also, $\mathbb{E}[\cdot]$ and $|\cdot|$ denote the
expected value and absolute value operators, respectively. Small letters and bold small letters show scalars and vectors, respectively. Further, $\norm{\mathbf{x}}_1=\sum_i |x_i|$ denotes the 1-norm of vector $\mathbf{x}$ and superscript $\left(\cdot\right)^T$ denotes the transpose operator.
\section{Background}\label{sec:RAMA}
\begin{figure}
\includegraphics[scale = 0.9]{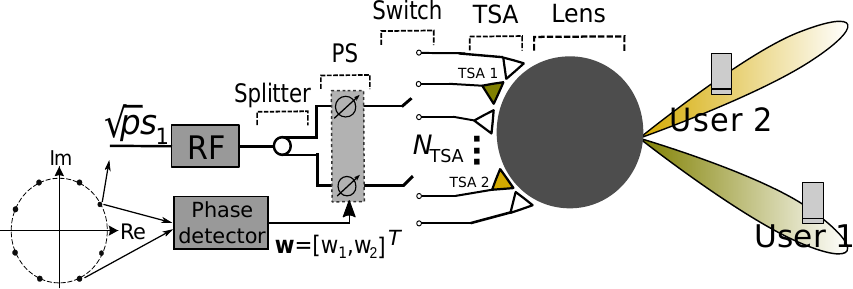}
\centering
\caption{\footnotesize Schematic of RAMA for two users, $\mathbf{w}=\left[ w_1, w_2 \right]^T = \left[ 1, e^{j{\Delta \theta}}\right]^T$~\cite{almasi2018reconfigurable}.}
 \label{fig2}
\end{figure}

We consider the downlink transmission in a single-cell mmWave communication system with a base station (BS) serving multiple users. The BS is equipped with a lens-based reconfigurable antenna and the users are provided with a single antenna. The reconfigurable antenna comprises of following blocks: one RF chain, a splitter, a PS, a switching network, $N_{\text{TSA}}$ tapered slot antennas (TSAs), and a spherical lens located in front of the TSAs~\cite{r19}. The system is shown in Fig.~\ref{fig2}. The scheme in Fig.~\ref{fig2} which uses one RF chain is based on the design in~\cite{almasi2018reconfigurable}, and we refer to the system as RAMA. 
The mmWave propagation environment is modeled by widely adopted multipath Saleh-Valenzuela channel model~\cite{el2014spatially,r9}. Suppose only one propagation channel is selected from the transmitter to each user. So, the channel for User $i$ is given by $\mathbf{h}_i=h_i\mathbf{a}(\theta_i,\phi_i)$ where $h_i \in \mathbb{C}$ denotes the channel gain of User $i$, $\mathbf{a}(\theta_i,\phi_i)$ denotes the antenna array response vector of the BS, and $\theta_i$ ($\phi_i$) $\in~[0, 2\pi]$ denotes the azimuth (elevation) AoD. The array vector is defined as
\begin{equation}\label{array}
\Scale[0.95]{\mathbf{a}(\theta,\phi) = \frac{1}{\sqrt{N_\text{ray}}}\left[1,\dots,e^{-j\pi\psi_{r,s}}, \dots, e^{-j\pi\psi_{N_{\text{ray},x}-1,N_{\text{ray},y}-1}}\right]^T},
\end{equation}
where $N_\text{ray}=N_{\text{ray},x}N_{\text{ray},y}$ and $\psi_{r,s}=\frac{2d_0}{\lambda}\big(r\text{sin}\theta\text{cos}\phi+s\text{sin}\theta\text{sin}\phi\big)$ for $r\in \{0, 1, \dots, N_{\text{ray},x}-1\}$ and $s\in \{0, 1, \dots, N_{\text{ray},y}-1\}$. $N_{\text{ray},x}$ and $N_{\text{ray},y}$ denote the number of rays of $x$ axis and $y$ axis, respectively. Also, $d_0$ and $\lambda$ are the antenna spacing and the wavelength, respectively.

In what follows, we briefly describe RAMA. In RAMA, we transmit only the intended signal for each user at the same time/frequency/code blocks. Assume only two users are available. The intended signals for Users $1$ and 2 are denoted by $s_1$ and $s_2$, respectively. Assume that $s_i$, for $i = 1,~2$, is drawn from a phase shift keying (PSK) constellation\footnote{Notice that while RAMA is compatible with all standard constellations, throughout this paper, we consider only PSK constellation. For more details on the other constellations, the interested reader is referred to~\cite{almasi2018reconfigurable}.} and $\mathbb{E}[|s_i|^2] = 1$. Then, $\mathbf{s}$ can be expressed in terms of $s_1$ as
\begin{align}\label{eq5}
\mathbf{s} = \mathbf{w}s_1,
\end{align}
where $\mathbf{s}=\left[s_1, s_2\right]^T$ and $\mathbf{w}=\left[1, e^{j{\Delta \theta}}\right]^T$ where  $\Delta\theta$ denotes the difference between the phases of $s_1$ and $s_2$. This module is implemented by a phase detector block shown in Fig.~\ref{fig2}. In the above design, only one of the signals, say $s_1$, is upconverted by the RF chain block and the whole power $p$ is allocated to that signal. 
The phase detector block calculates the phase difference between $s_1$ and $s_2$, i.e., $\Delta \theta$. This block is shown as a black box in Fig.~\ref{fig2}. The switch network selects two TSA feeds, e.g., the green (TSA1) and yellow (TSA2) feeds in Fig.~\ref{fig2}. The signals corresponding to TSAs 1 and 2 are given by $\sqrt{p_1}s_1$ and $\sqrt{p_2}s_2$, respectively. 
\begin{figure*}[h]
    \centering
    \includegraphics[scale=0.70]{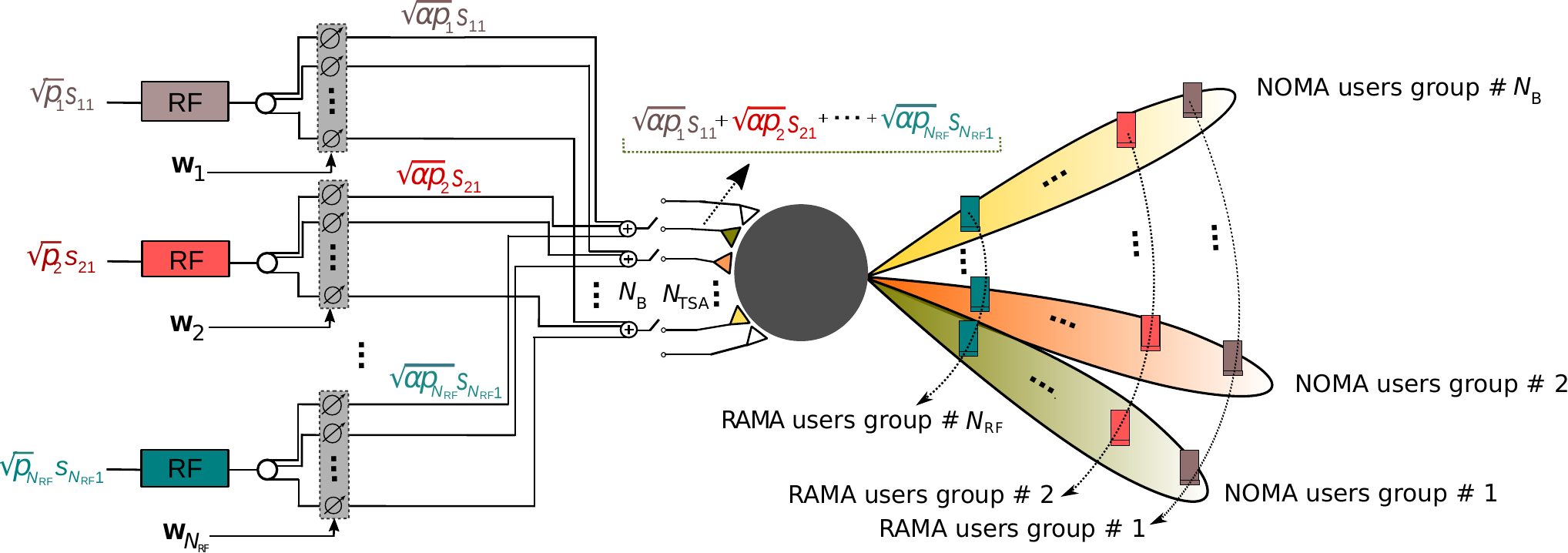}
    \caption{The proposed RA-NOMA technique. The users are grouped into RAMA users and NOMA users. The number of RAMA groups is equal to the number of RF chains and the number of NOMA groups is the same as the number of beams. The number of beams is independent of the number of RF chains.}
    \label{fig:RA-NOMA}
\end{figure*}

As illustrated in Fig.~\ref{fig2}, in the reconfigurable antenna, a spherical electromagnetic lens is positioned in front of the antenna array (TSA feeds). The lens operates as a passive PS network that shifts the phase of the signal with different delays such that when a plane wave like $h_i\mathbf{a}(\theta_i, \phi_i)$ hits the lens from one side, the output on the other side is only the magnitude of the wave $h_i$~\cite{abbasi2018lens}. Since transmission is directional in mmWave bands, each RAMA user receives only its intended signal, $y_i$ for $i$=1, 2, i.e.,
\begin{align}\label{eq6}
\begin{cases}
y_1 = \sqrt{p_1}h_1s_1 + n_1,\\
y_2 = \sqrt{p_2}h_2s_2 + n_2,\\
\end{cases}
\end{align}
where $n_i$  denotes the noise term which is additive white Gaussian with zero-mean and variance $\sigma_n^2$. It is worth mentioning that the lens constructively delays the phase of elements of the antenna array response vector, however, the PSs change the phase of the $s_i$ in a way to obtain the desired signals, i.e., $\mathbf{s}= [s_1, s_2]^T$. Here, the splitter routes input power equally between its outputs, i.e., $p_1 = p_2 = \frac{p}{2}$. This is due to the fact that equally-divided power splitters can reduce hardware complexity compared to the dynamic counterpart. The achievable rate of RAMA for each user under equal power allocation ($p_1=p_2 =\frac{p}{2}$) is obtained as~\cite{almasi2018reconfigurable} 
\begin{align}\label{eq7}
\begin{cases}
R_1 = \text{log}_2(1 + \frac{p_1|h_1|^2}{\sigma_n^2}), \\
R_2 = \text{log}_2(1 + \frac{p_2|h_2|^2}{\sigma_n^2}).
\end{cases}
\end{align}
In~\cite{almasi2018reconfigurable} it is shown that when users experience different AoDs (i.e., transmission paths), RAMA is a more efficient than NOMA to serve multiple users.  

It is noteworthy that, here, we assume that no interference is imposed from the signal intended for User 1 on User 2 and vice versa. This assumption is very well justified since the proposed lens based slotted reconfigurable antenna results in highly directional beams with very limited sidelobes~\cite{schoenlinner2002wide}. Moreover, due to significant path loss and shadowing in mmWave frequencies we do not expect the signals from the sidelobes to reach the unintended user\footnote{Detail analysis of the impact of sidelobes on inter-beam interference in RAMA is subject of future research.}. 

\section{The Proposed RA-NOMA Technique}\label{sec:RA_NOMA}

Recall that RAMA is only applicable to users with different AoDs and in directional transmission whereas NOMA is mostly applied for users with the same AoDs but different channel gains. This points out that the two techniques can be used to serve two different groups of users. Hence, there is a unique opportunity to integrate RAMA with NOMA, i.e., RA-NOMA, to add a degree of freedom to serve far more users without increasing the number of RF chains.

Toward this goal, we divide the users into two groups based on their AoDs and channel gains as: \rmnum{1}) RAMA and \rmnum{2}) NOMA groups. The users in the RAMA group are assumed to have distinctive AoDs but relatively similar channel gains. The users within the NOMA group are assumed to have the same AoDs but different channel gains. Note that probability of having users with the same AoDs and same
channel gains is small. However, the probability of having users with similar AoDs or similar channel gains is high in 5G networks due to dense deployment~\cite{7422408}. Fig.~\ref{fig:RA-NOMA} depicts a potential configuration for RA-NOMA users. The users with the same color belong to the same RAMA group. Further, those users covered by the same beam belong to the same NOMA group.

In our proposed design, we assume that the splitters have a fixed number of outputs with equal power. Therefore, the number of beams supporting the NOMA groups is fixed. We assume that the number of beams, i.e., the NOMA groups, is $N_B$. The number of RAMA groups can vary up to the number of RF chains. Therefore, $N_\text{RF}$ is the number of users supported by each beam. Hence, the total number of the served users via RA-NOMA is $N_B \times N_\text{RF}$.

The transmission power of the $i$th RF chain, $i=1,\dots,N_{\text{RF}}$, i.e., total allocated power to the $i$th group of RAMA users, is assumed to be $p_i$. Also, $\alpha p_i$ denotes the allocated power to each user in the $i$th RAMA group in which $\alpha =\frac{1}{N_B}$. 
We let $s_{i1}$ denote the intended signal for the $i$th group of RAMA users before the $i$th RF chain. At the $i$th PS, the vector $\mathbf{w}_i$ is designed based on~\eqref{eq5}, such that we have
\vspace{-5pt}
\begin{equation}
    \mathbf{w}_i = [1, e^{j\Delta\theta_{i2}}, \dots, e^{j\Delta\theta_{iN_B}}]^T,
\end{equation}
and
\vspace{-10pt}
\begin{align}
\mathbf{s}_i=\mathbf{w}_i s_{i1},
\end{align}
where, $\mathbf{s}_i=[s_{i1},\dots, s_{iN_B}]^T$ contains the intended signals for the users in the $i$th RAMA group, and $\Delta\theta_{ik},~k=2,\dots,N_B,$ denotes the difference between the phases of $s_1$ and $s_k$ in the $i$th RAMA group. Considering the allocated power of each NOMA group, the superposition coded signal of the $k$th beam is given by
\begin{equation}
    \sum_{i=1}^{N_\text{RF}} \sqrt{\alpha p_i}s_{ik} = \sqrt{\alpha p_1}s_{1k}+\dots+\sqrt{\alpha p_{N_\text{RF}}}s_{N_\text{RF}k}.
\end{equation}

Let $h_{ik}$ refer to the channel gain of a user in the $i$th RAMA group in the $k$th beam and we refer to this user as User $\left(i,k\right)$. In the $k$th beam or equally the $k$th NOMA group, without loss of generality, we assume the channels of users are sorted as $|h_{1k}|^2 \leq |h_{2k}|^2 \leq \dots \leq |h_{N_\text{RF} k}|^2$. 
The received signal by User $\left(i,k\right)$ becomes
\vspace{-5pt}
\begin{align}
    y_{ik}=\underbrace{\sqrt{\alpha p_i}h_{ik}s_{ik}}_{\text{intended signal}} + \underbrace{\sum_{l=1, l \neq i}^{N_\text{RF}}\sqrt{\alpha p_l}h_{ik}s_{lk}}_{\text{intra-beam interference}}+\underbrace{n_{ik}}_{\text{noise}},
\end{align}
where $n_{ik}$ is the additive white Gaussian noise with variance $\sigma^2$. As it is seen, inter-beam interference is removed. In each beam, with using NOMA, successive detection is carried out in descending order. Therefore, the achievable rate of User $\left(i,k\right)$ is obtained as
\vspace{-5pt}
\begin{equation}\label{eq_rate}
    R_{ik}=\text{log}_2\left(1 + \frac{\alpha p_i|h_{ik}|^2}{|h_{ik}|^2 \sum_{l= i+1}^{N_\text{RF}}\alpha p_l + \sigma^2}\right).
\end{equation}

\section{Power Allocation For Downlink RA-NOMA}\label{sec:power}

In this section, we study power allocation for the proposed RA-NOMA system. Let $\mathbf{p}=\left[p_1, \dots, p_{N_\text{RF}} \right]^T$ be the transmission powers of the RAMA groups. Our objective is to optimize the power allocation to maximize the sum achievable rate under the total power and individual minimum rate constraints for users as follows
\begin{maxi!}[2]
{\mathbf{p}}{ \sum_{k=1}^{N_B} \sum_{i=1}^{N_{\text{RF}}} R_{ik}\label{eq_objectiveopt}}{\label{eq_opt}}{}
\addConstraint{\sum_{i=1}^{N_{\text{RF}}} p_i}{\leq P_{\max}\label{a}}
\addConstraint{R_{ik}}{\geq \bar{R}_{ik},\label{b}}{\forall i,k}
\addConstraint{\mathbf{p}}{\geq \mathbf{0},\label{c}}
\end{maxi!}
where, $P_{\max}$ is the maximum transmit power of the system and $\bar{R}_{ik}$ denotes the minimum (required) rate at User $\left(i,k\right)$. The defined power allocation problem above can ensure user fairness by assuming the same minimum rate requirement for all users in the constraint~\eqref{b}.
\begin{lemma}\label{lemma1}
The maximum of minimum rate requirement in the $i$th RAMA group defines the minimum power requirement for the $i$th RF chain.
\begin{proof}
See Appendix~\ref{appen_Lemma1}.
\end{proof}
\end{lemma}
Using Lemma~\ref{lemma1}, it is easy to show that each beam achieves the same transmission rate. Therefore, the power optimization problem in~\eqref{eq_opt} is transformed into the power allocation over one beam. We define $\bar{R}_i=\underset{k}\max~{\bar{R}_{ik}}$, $|h_i|=|h_{i,k}|$, for $k \in \left\lbrace 1,\dots,N_B\right\rbrace$, and $a_i = \sum_{l=i}^{N_{\text{RF}}} p_l$. Considering the above, the optimization problem over one beam is as follows.
\begin{maxi!}[2]
{\mathbf{p}}{ \sum_{i=1}^{N_{\text{RF}}} \text{log}_2\left(\frac{|h_{i}|^2 a_i + \frac{\sigma^2}{\alpha}}{|h_{i}|^2 a_{i+1}+ \frac{\sigma^2}{\alpha}}\right) \label{eq_objectiveopt2}}{\label{eq_opt2}}{}
\addConstraint{\sum_{i=1}^{N_{\text{RF}}} p_i}{\leq P_{\max}\label{a2}}
\addConstraint{|h_{i}|^2 a_i + \sigma^2 / \alpha}{\geq 2^{\bar{R}_i}\left(|h_{i}|^2 a_{i+1}+ \sigma^2 / \alpha \right),\label{b2}}{\forall i}
\addConstraint{\mathbf{p}}{\geq \mathbf{0}.\label{c2}}
\end{maxi!}
The optimization problem in~\eqref{eq_opt2} has a feasible solution if there is any distribution of $\mathbf{p}$ such that satisfies the minimum rate requirements while meeting the constraint in~\eqref{a2}. Therefore, we consider minimizing the total transmission power with rate constraints for users as 
\begin{mini!}
{\mathbf{p}}{ \norm{\mathbf{p}}_1 \label{eq_objectiveopt3}}{\label{eq_opt3}}{}
\addConstraint{R_{i}}{\geq \bar{R}_{i},}{\forall i.}{\label{b3}}
\end{mini!}
We define the answer to the optimization problem in~\eqref{eq_opt3} as $\mathbf{p^*}$ and the feasibility condition of problem~\eqref{eq_opt2} becomes $\norm{\mathbf{p^*}}_1 \leq P_{\max}$. Due to lack of space, we use the results in~\cite{art_jinho} for solving~\eqref{eq_opt3} and the optimal power for the $p_i^*$ can be derived as
\begin{align}
p_i^*=\left(\sum_{l=i+1}^{N_\text{RF}} p_{l}^* + \frac{\sigma^2}{\alpha |h_{l}|^2} \right) \left( 2^{\bar{R}_{i}} -1  \right).
\end{align}
Therefore, the optimal value of~\eqref{eq_opt3} is $\norm{\mathbf{p^*}}_1=\sum_{i=1}^{N_{\text{RF}}} p_i^*$. By some manipulations we have 
\begin{align}\label{eq_feasible}
\norm{\mathbf{p^*}}_1=\sum_{l=1}^{N_{\text{RF}}} \frac{\left(\Pi_{m=1}^{l-1} 2^{\bar{R}_m}\right) \left(2^{\bar{R}_l}-1\right) \frac{\sigma^2}{\alpha}}{|h_l|^2} \leq P_{\max},
\end{align}
as the feasibility condition of the optimization problem in~\eqref{eq_opt2}.

The objective function in~\eqref{eq_objectiveopt2} is proved to be concave in~\cite{art_convexity}. Since the constraints in~\eqref{eq_opt2} are linear, the optimization problem is convex. Therefore, the solution of the problem can be found by solving the Karush-Kuhn-Tucker (KKT) conditions~\cite{Boyd}. The Lagrangian function of the problem~\eqref{eq_opt2} is given by
\begin{equation}\label{eq_cond2}
\begin{aligned}
\mathcal{L} & \left(\mathbf{p},  \gamma, \pmb{ \beta} \right)= \sum_{i=1}^{N_{\text{RF}}} \text{log}_2\left(\frac{|h_{i}|^2 a_i + \frac{\sigma^2}{\alpha}}{|h_{i}|^2 a_{i+1}+ \frac{\sigma^2}{\alpha}}\right) \\ & + \gamma \left(P_{\max}-\sum_{i=1}^{N_{\text{RF}}} p_i\right) \\ & + \sum_{i=1}^{N_{\text{RF}}} \beta_i \left(|h_{i}|^2 a_i + \frac{\sigma^2}{\alpha} - 2^{\bar{R}_i}\left(|h_{i}|^2 a_{i+1}+ \frac{\sigma^2}{\alpha} \right) \right),
\end{aligned}
\end{equation}
where, $\gamma$ and $\pmb{\beta}=\left[\beta_1,\dots,\beta_{N_{\text{RF}}}\right]^T$ are the non-negative Lagrangian multipliers associated with~\eqref{a2} and~\eqref{b2}, respectively. From this point, the KKT conditions and solution is the same as~\cite{art_convexity}. Therefore, to save space and avoid replication, we just state the results of the solution as follows
\begin{itemize}[leftmargin=*]
\item The Lagrange multiplier $\gamma$ is greater than zero and constraint~\eqref{a2} holds with equality.
\item The Lagrange multipliers $\beta_i$ for $i=1,\dots,N_{\text{RF}-1}$ are greater than zero and the solution of the problem depends on the multiplier $\beta_{N_{\text{RF}}}$.
\item If $\beta_{N_{\text{RF}}} > 0$, the constraint~\eqref{b2} holds with equality for $i=1,\dots,N_{\text{RF}}$ and the optimal value of problem~\eqref{eq_opt2} (maximum achievable rate for one beam) is $\sum_{i=1}^{N_{\text{RF}}} \bar{R}_i$ and 
if $\beta_{N_{\text{RF}}}=0$, the optimal value is as follows
\begin{equation}
\text{log}_2 \Scale[0.95]{\left( 1 + \frac{\alpha P_{\max} |h_{N_{\text{RF}}}|^2}{\sigma^2 \Pi_{i=1}^{N_{\text{RF}}-1} 2^{\bar{R}_i}} - \sum_{i=1}^{N_{\text{RF}}-1} \frac{|h_{N_{\text{RF}}}|^2 \left( 2^{\bar{R}_i} - 1\right)}{|h_i|^2 \Pi_{l=i}^{N_{\text{RF}}-1} 2^{\bar{R}_l}} \right)}+ \sum_{i=1}^{N_{\text{RF}}-1} \Scale[0.95]{\bar{R}_i}.
\end{equation}
\end{itemize}
Consequently, the optimal value of problem~\eqref{eq_opt} is that of problem~\eqref{eq_opt2} multiplied by $N_B$. 

\section{Simulation Results}\label{sec:sim}

In this section we evaluate the performance of the proposed RA-NOMA by using numerical analysis. We consider $12$ users to be served by RA-NOMA. The users are scheduled in three RAMA groups and four beams as shown in Fig.~\ref{fig:RA-NOMA}. The minimum rate requirements are considered to be equal and set to $\bar{R}_{i,k}=0.2$ (b/s/Hz). We define transmit signal-to-noise ratio (SNR) as the normalized transmit power with respect to $\sigma^2$. 
We assume that the users in RAMA groups have $|h|^2/\sigma^2=\left\lbrace 0, -5, -10\right\rbrace$ dB.

\begin{figure}
\vspace*{-0.4cm}
    \centering
    \includegraphics[scale=0.5]{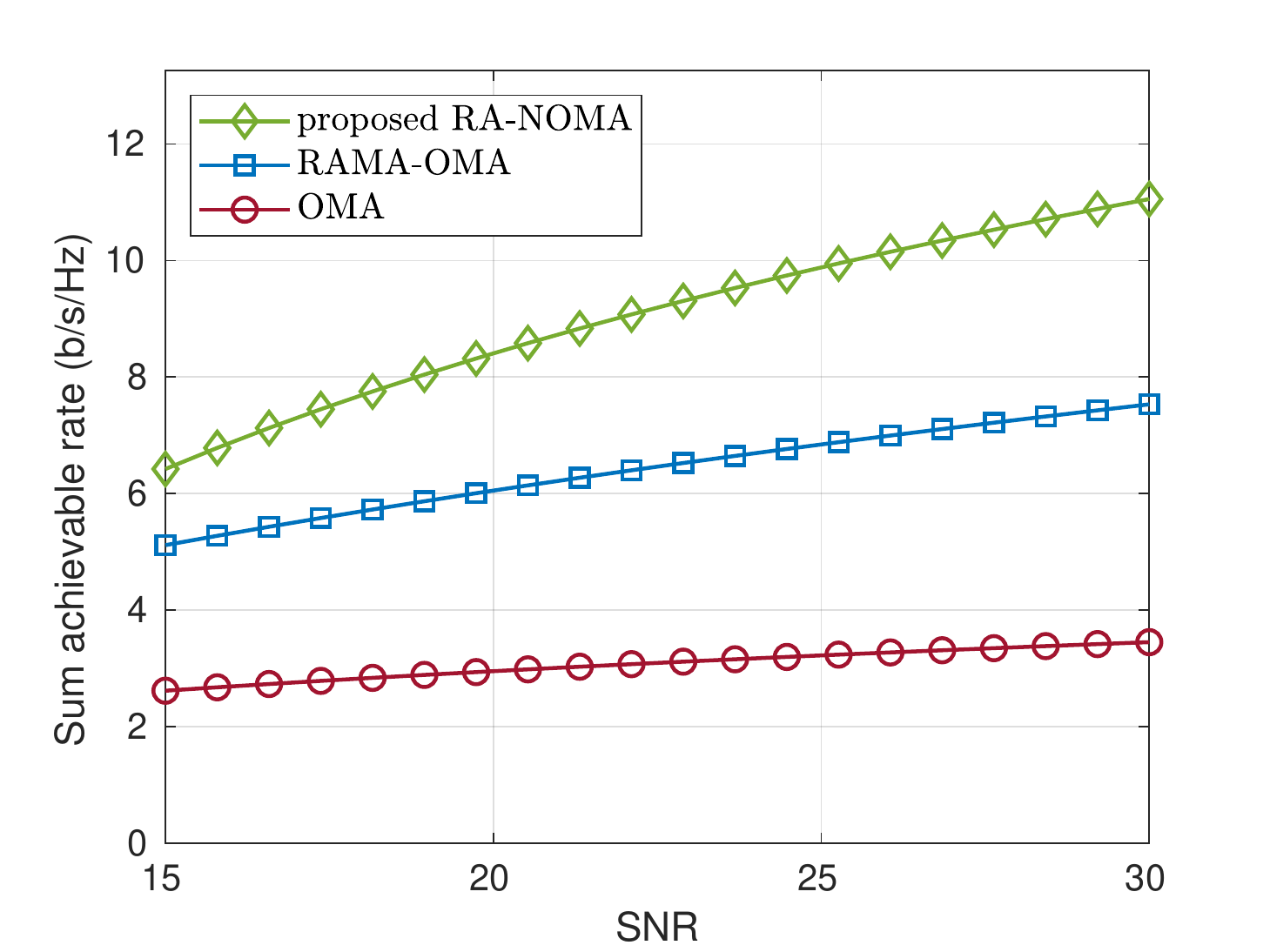}
    \caption{\footnotesize Sum achievable rate of OMA, RAMA-OMA, and RA-NOMA with respect to transmit SNR.}
    \label{fig:sum_rate}
\end{figure}

The performance of RA-NOMA is compared with two access methods: OMA and RAMA-OMA. Table~\ref{table_PL} provides the number of required RF chains and time slots for the mentioned techniques. 
Fig.~\ref{fig:sum_rate} represents sum-rate of the three techniques versus the transmit SNR. 
The simulation results are normalized for one time slot for all three techniques. As it is shown, the proposed RA-NOMA outperforms OMA and RAMA-OMA techniques in terms of sum-rate with equal power budget at the transmitter. It should be noticed that RA-NOMA is not compared with NOMA technique. This is because user grouping regarding Fig.~\ref{fig:RA-NOMA} may not be appropriate for deploying NOMA technique. As such, NOMA technique for this grouping requires four RF chains which is not efficient in terms of energy consumption and hardware expenses. One would deploy NOMA with lens antenna and three RF chains for $12$ users. The grouping should contain three beams and supports four users per each beam. However, due to lack of space, the comparison between RA-NOMA and NOMA with different user groupings will be provided in the extended version of this article.

\begin{table}
\centering
\setlength\doublerulesep{1.0pt}
\caption{Number of required RF chains and time slots for each technique.}
\label{table_PL}
\begin{tabular}{|c|c|c|c|} \hline
    & \textbf{OMA}		&\textbf{RAMA-OMA}		   & \textbf{RA-NOMA} \\ \hline
	Number of RF chains			& 1  &1  &3 \\ \hline
	Number of time slots        & 12  &3  &1 \\        \hline
\end{tabular}
\end{table}

\section{Conclusion and Future Work}\label{sec:conclusion}
In this paper, we proposed  RA-NOMA as a new multiple access technique
for mmWave lens-based reconfigurable antennas. To achieve this, we  take advantage of both RAMA and NOMA techniques. Due to the directive and independent beams steered by the lens antennas, the inter-cluster interference between the users is eliminated. Further, the proposed RA-NOMA simultaneously supports a large number of users by using a single BS in the downlink and less number of RF chains compared to the existing mmWave-NOMA technique. 
The maximum achievable downlink rate of RA-NOMA is derived and the simulation results demonstrate that the proposed RA-NOMA achieves higher sum-rate compared to RAMA and conventional OMA techniques. For future work, the authors will investigate the application of learning techniques in clustering and power allocation~\cite{art_Amiri_ICC18} of the proposed RA-NOMA technique.
\vspace{-10pt}
\appendices
\section{}\label{appen_Lemma1}
The minimum rate requirement constraint in~\eqref{b} imposes the following condition for $p_i$.
{\begin{align}
p_i \geq \underbrace{\left(\sum_{l=i+1}^{N_\text{RF}} p_{l} + \frac{\sigma^2}{\alpha |h_{ik}|^2} \right)}_{\left(a\right)} \underbrace{\left( 2^{\bar{R}_{ik}} -1  \right)}_{\left(b\right)}.
\end{align}}
We recall that users of the $i$th RAMA group are scheduled with almost equal channel gains, i.e, $|h_{i,k}|\approx |h_{i,k^\prime}|$, for $k,k^\prime \in \left\lbrace 1,\dots,N_B\right\rbrace$. Therefore, the term $\left(a\right)$ has almost the same value for all users in one RAMA group. So, the minimum power of the $i$th RAMA group, $i$th RF chain, is lower bounded by the maximum of the term $\left(b\right)$, i.e., $2^{\underset{k}\max~{\bar{R}_{ik}}} -1$.
\bibliographystyle{IEEEtran}
\bibliography{IEEEabrv,references}
\end{document}